# Non-equilibrium hot-carrier transport in type-II multiple-quantum wells for solar-cell applications

H. P. Piyathilaka,[1] R. Sooriyagoda,[1] V. R. Whiteside,[2] T. D. Mishima,[2]
M. B. Santos,[2] I. R. Sellers[2] and A. D. Bristow[1,*]

[1] *Department of Physics and Astronomy, West Virginia University, Morgantown, West Virginia 26501-6315, USA*
[2] *Department of Physics and Astronomy, University of Oklahoma, Norman, Oklahoma, 73019, USA*
*E-mail: alan.bristow@mail.wvu.edu*

**Abstract:** Prototypes for hot-carrier solar cells based on type-II InAs/AlAsSb multiple quantum wells are examined for AC photoconductivity as a function of lattice temperature and photoexcitation energy to determine the photoexcited charge carrier transport. These samples previously exhibit an excitation energy onset of a metastable regime in their short time charge carrier dynamics that potentially improves their applicability for hot-carrier photovoltaic applications. The transport results illustrate that the AC photoconductivity is larger in the dynamic regime corresponding to the metastability as a result of higher excitation photocarrier densities. In this excitation regime, the AC photoconductivity is accompanied by slightly lower carrier mobility, arising from the plasma-like nature of carriers scattered by Auger recombination. Outside of this regime, higher mobility is observed as a result of a lower excitation density that is more readily achievable by solar concentration. Additionally, at ambient temperatures, more scattering events are accompanied by slightly lower mobility, but the excitation dependence indicates that this is accompanied by an ambipolar diffusion length that is greater than half a micron. These transport properties are consistent with good quality in-organic elemental and III-V semiconductor solar cells and far exceed those of novel materials. The transport results complement the dynamics observed in type-II InAs/AlAsSb and can guide the design of hot-carrier solar cells based on these and related materials.

## 1. Introduction

Hot-carrier solar cells (HCSCs) have been proposed to overcome the traditional single-junction detailed-balanced limit.[1–4] In conventional solar cells, broad-spectrum light photoexcites many carriers with excess energy above the band gap and will undergo thermalization through cooling and relaxation.[4,5] This process converts photonic energy to thermal energy, rather than usable electrical energy. In HCSC designs, excited carriers should be extracted before they thermalize and cool, and thus overcome thermodynamic limits by reducing the energy loss to heat.[3,6–8] To achieve this goal, carrier recombination should be slow. Silicon solar cells exploit indirect recombination, which can be replicated in type-II aligned multiple quantum well (MQW) and superlattice (SL) heterostructures. These structures can also create optical resonances to improve the desired absorption. Moreover, MQW/SL structures can have reduced thermal conductivity that further slows carrier cooling via phonon emission and subsequent phonon-phonon relaxation.[9–11]

Design of MQW/SL structures for HCSC applications requires significant electronic and phononic band gap engineering, which is available in heterostructures such as InAs/AlAsSb.[12] These systems exhibit type-II electronic band alignment, where conduction-band electrons are predominantly confined to the InAs wells and valence-band holes to the AlAs$_{0.16}$Sb$_{0.84}$ barriers. It has been observed that they have long-lived photoluminescence over a wide range of lattice temperatures,[12] a phonon-bottleneck that prevents rapid carrier cooling and even hot-carrier stabilization,[10,12,13] charge-carrier dynamics that show Auger scattering on short timescales, and a metastable state that prolongs hot-carrier lifetimes.[13] This metastability has an energy onset that may make it compatible with proposals for hot-carrier extraction through energy-selective contacts. Furthermore, devices with applied biases have shown increased charge separation by intervalley scattering that is expected to further slow electrons-holes recombination.[14–16]

Important parameters for designing a HCSC device include the optical absorption, hot-carrier lifetime and excited-state mobility. It is unknown whether the benefits observed in the photoexcited dynamics of these type-II MQWs apply to the excited-state charge-carrier transport, especially because alloy intermixing at the internal interfaces may reduce the carrier mobility.[10] In this paper, the transport of the an InAs/AlAs$_{0.16}$Sb$_{0.84}$ MQW structure is evaluated by AC-photoconductivity for a range of lattice temperatures and optical-excitation conditions. This is achieved using femtosecond optical pump pulses and terahertz (THz) probe pulses. This approach allows for measurements of the heterostructure without contact effects that require additional modelling parameters,[17] offers access to the complex conductivity of excited carriers[18–20] through time-domain spectroscopy,[21,22] and allows the transport to be tracked during the subsequent decay dynamics.[10,13,19,20] Measurements are performed in the previously observed metastable regime to determine if the carrier transport is adversely affected by the high excitation conditions. Fitting with a plasmon model yields carrier density and mobility. Temperature-dependent mobility indicates ambipolar diffusion. Whereas excitation-energy dependence suggests that transition into the metastability regime may be associated with transition into plasma-like transport suitable for strong Auger scattering.[13]

## 2. Experimental

A type-II InAs/AlAs$_{0.16}$Sb$_{0.84}$ MQW structure is grown using molecular beam epitaxy[23] to have 30 periods of 2.4 nm InAs wells with 10 nm of AlAs$_{0.16}$Sb$_{0.84}$ barriers. A 2-µm thick InAs buffer layer was grown on the GaAs substrate to relax the strain in order to produce a strain-relaxed heterostructure.[12] The substrate was removed using mechanical polishing and wet etching leaving the structure attached to a c-cut sapphire transfer substrate for optical measurements.[10] The inset of Figure 1(a) shows a schematic diagram of the sample structure. This structure has an optical absorption edge (band gap) of 0.857 eV at 4 K, which is non-monotonic with increasing temperature as various defect states become optically inactive and bands relax.[10,12]

Figure 1(a) shows the time-resolved THz spectroscopy setup using a 1-kHz regenerate laser amplifier to produce ~100-fs pulses centered at 800 nm.[13] Output pulses from the amplifier are split into two replicas, one for generation and detection of THz radiation, the other for optical excitation. The THz is created by optical rectification in a CdSiP$_2$ crystal[21] and detected by electro-optic (EO) sampling in a ZnTe crystal.[24–26] THz pulses propagate in a dry-air purged environment through four off-axis parabolic mirrors with an intermediate focus where the sample resides in a cryostat with temperature control from 4 K to 300 K. The pump photon energy is tuned in an optical parametric amplifier (OPA) with a range from 0.775 eV (1.6 µm) to 1.03 eV (1.2 µm).

The (excited-state) AC photoconductivity spectra are determined by the differential transmission of THz radiation through the sample with and without the optical excitation. Both time-domain signals are acquired by varying the THz delay time ($t_{THz}$), for a fixed pump-probe delay time ($\Delta t$). A mechanical chopper is placed in the THz generation beam for reference spectrum (without pumping) and in the optical excitation beam for the differential measurement and in both cases the chopper's modulation frequency is fed into the lock-in amplifier to read out the differential electro-optic signal. Figure 1(b) shows examples of the THz transients from the substrate (orange symbols), sample (black symbols) and photo-excited sample (red symbols), showing the main feature around $t = 0$ and a reflection in the detection crystal at $t_{THz} \sim 10$ ps. Spectra are obtained by performing a numerical Fourier transform of the respective transients with a window function to remove the unwanted reflection; see inset of Fig. 1(b). The maximum optical excitation density was maintained at ~$10^{13}$cm$^{-2}$ per well for all lattice temperatures ($T_l$), pump-probe delay times and excitation photon energies.

Complex AC-photoconductivity spectra $\Delta\tilde{\sigma}(\omega)$ is determined from the time-domain THz transmission spectra for the ground $\tilde{T}(\omega)$ and excited states $\Delta\tilde{T}(\omega)$, given by[19]

$$\Delta\tilde{\sigma}(\omega) = -\frac{\epsilon c(1 + \tilde{n}_s)}{d}\frac{\Delta\tilde{T}(\omega)}{\tilde{T}(\omega)}, \quad (1)$$

where $\tilde{n}_s$ is the refractive index of the substrate – measured as a function of temperature and consistent with literature,[27,28] $d$ is the sample thickness, $\epsilon_0$ and $c$ are the vacuum permittivity and light speed. The ground- and excited-state transmission spectra are determined from the ratios $\tilde{T}(\omega) = E_{gs}(\omega)/E_{sub}(\omega)$ and $\Delta\tilde{T}(\omega) = \Delta E_{es}(\omega)/E_{sub}(\omega)$, where $E_{gs}(\omega)$, $\Delta E_{es}(\omega)$ and $E_{sub}(\omega)$ are the ground-state, excited-state and substrate THz field spectra respectively, all measured as a function of temperature.

## 3. Results

Figure 2 (a) show the real (solid symbols) and imaginary (open symbols) parts of $\Delta\tilde{\sigma}(\omega) = \sigma_r(\omega) + i\sigma_i(\omega)$ for a series

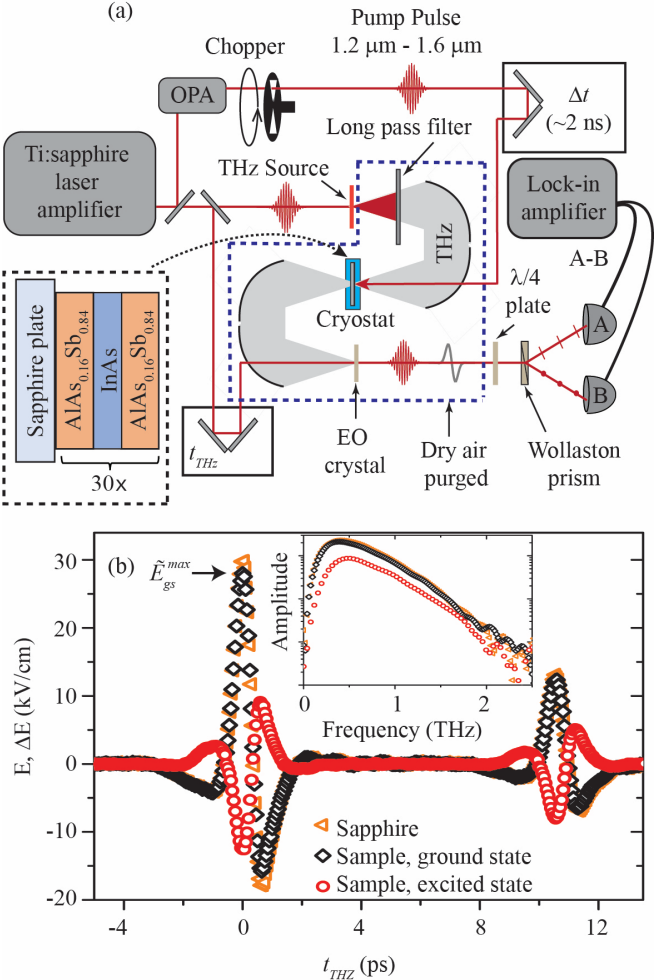

**Fig. 1** (a) Time-resolved terahertz spectroscopy setup and sample structure inset. (b) Transient THz transmission through the sapphire substrate, the InAs/AlAs$_{0.16}$Sb$_{0.84}$ MQW without photoexcitation, and the differential transmission trough the sample with photoexcitation. The inset shows the amplitude spectra from numerical Fourier transforms.

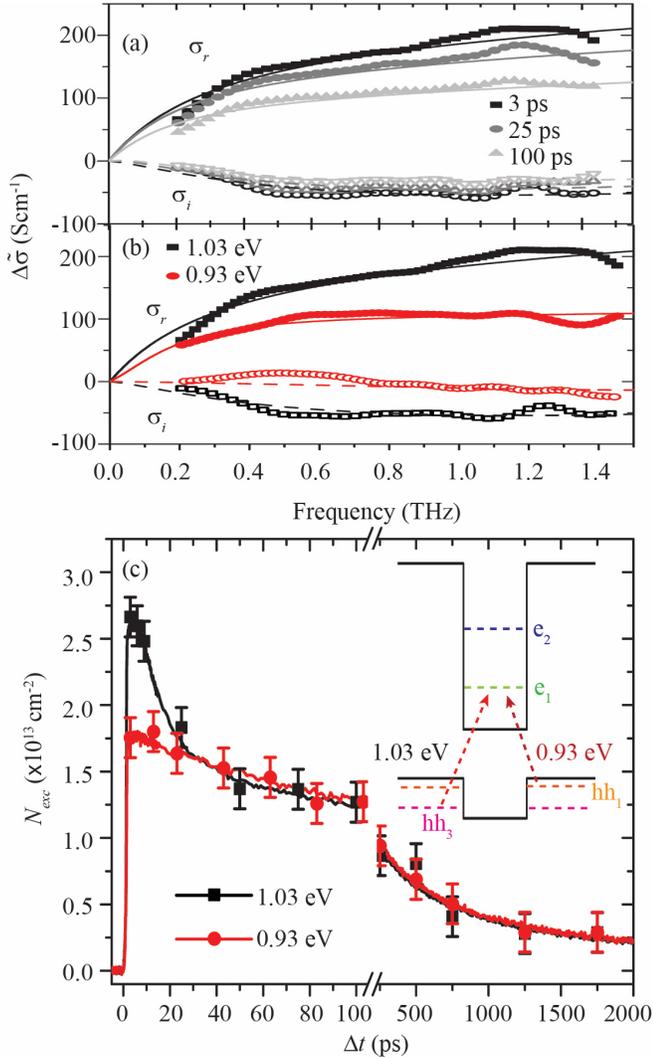

**Fig. 2** AC photoconductivity spectra for the MQW sample at a lattice temperature of 4 K for (a) a range of pump-probe delay times with excitation of 1.03 eV, and (b) excitation of 0.93 eV and 1.03 eV for $\Delta t = 3$ ps. Results fit to a plasmon conductivity model. (c) Photocarrier density $N_{exc}$ versus pump-probe delay time measured by transient absorption of $\tilde{E}_{gs}^{max}$ (solid lines), overlaid with values from the plasmon model (data points). Inset shows the excitation schemes in a single quantum well.

of pump-probe delay times with fixed excitation at 1.03 eV and lattice temperature of 4 K. Increasing $\Delta t$ reduces the excited-carrier density $N_{exc}$, because carriers relax and recombine by various mechanisms.[29,30] Conductivity is directly proportional to $N_{exc}$, so $\sigma_r$ (and to some extent $\sigma_i$) is observed to decrease with increasing $\Delta t$, as expected. However, as carrier density decreases, the mean-free path $l_f$ of those excited carriers will increase. The scattering time is $\tau = l_f/v_{th}$, where $v_{th} = \sqrt{3k_B T_l/m^*}$ is the thermal velocity of the carrier, with $k_B$ is the Boltzmann constant and $m^*$ is the effective mass. Estimated values for $l_f$ are presented in Table 1 and discussed later. Nevertheless, the dependence of $\tau$ on $l_f$ is indirectly dependent on $N_{exc}$, such that the conductivity $\Delta\tilde{\sigma}(\omega)$ may not decrease directly proportionally to $N_{exc}$.

Figure 2(b) shows $\sigma_r$ and $\sigma_i$ for 0.93 eV (red symbols) and 1.03 eV (black symbols) excitation, for $\Delta t = 3$ ps and $T_l = 4$ K. Higher excitation energy has a stronger conductivity, which is indicative of a higher $N_{exc}$. This is consistent with data determined from the photon-energy dependent excitation dependence and dynamics reported previously.[13] Figure 2(c) shows the transient absorption recorded at the maximum of the ground-state THz field $\tilde{E}_{gs}^{max}$ [see Fig. 1(b)] and converted into excited-carrier density by $N_{exc} = -\alpha d \log(\Delta\tilde{E}_{es}^2/\tilde{E}_{gs}^2)$, where the sample absorbance as $\alpha d \approx 0.19(0.24)$ for 0.93(1.03) eV excitation. Results confirm that at short delay times ($\Delta t < 30$ ps), $N_{exc}$ is indeed higher for the 1.03 eV excitation. These two photon energies straddle the onset of a metastable response, associated with an indirect hh3→e1 transition >0.98 eV and a higher density of states. The dynamics in this regime have shown an increased Auger-scattering during early times that prolongs hot carriers beyond their lifetime arising from the phonon bottleneck due to a reduced thermal conductivity.[10]

Spectra in Figs. 2(a) and (b) were fit to various conduction models.[20] Despite partial confinement of the carriers, to match positive $\sigma_r$ and negative $\sigma_i$, with both trending toward zero at low frequency, the best fit was a plasmon model given by

$$\Delta\tilde{\sigma}_P(\omega) = \left(\frac{N_{exc}e^2}{m^*}\right)\left(\frac{\tau}{1 - i\tau(\omega - \omega_0^2/\omega)}\right), \quad (2)$$

where $e$ is the elementary charge, $m^*$ is effective mass of InAs, and $\omega_0 = \sqrt{g(N_{exc}e^2/\varepsilon_0 m^*)}$ with $g = 0.52$ is a geometry factor.[31] It is found that $\omega_0$ changes linearly with respect to the $\sqrt{N_{exc}}$ from 0.95 THz to 2.55 THz. The plasmon model describes carriers driven by the external electromagnetic wave with a restoring force. Such a restoring force leads to damping of the charge-carrier transport and can be provided by a depletion or accumulation field in semiconductor nanostructures. Here the model agrees well with the measured spectra for all pump-probe delay times and for both photoexcitation energies.

Figure 2(c) shows the extracted $N_{exc}(\Delta t)$ (data points) overlaid on the directly measured transients. The transients only measure variations in the maximum $E_{gs}(\omega)$ in the THz trace while varying $\Delta t$ and are insensitive to subtleties of the full THz spectrum. On the other hand, fitting the full THz spectrum by the plasmon model considers sample geometry and carrier-scattering mechanisms, such that $N_{exc}$ can return values that do not match that of the transient. Good agreement between the two approaches further confirms the validity of the plasmon model and will provide confidence for determining the mobility as a function of carrier density in later discussions.

The temperature and excitation dependence of the excited-carrier transport in the metastable regime is obtained using the $N_{exc}$ can be extracted for lattice temperature and pump-probe delay time.

Figure 3(c) shows $N_{exc}(\Delta t)$ for 1.03 eV photoexcitation $T_l = \{4, 100, 200\ \&\ 300\}$ K determined from the transient absorption (solids lines) and the plasmon model (data points). The measured and extracted $N_{exc}$ agree well for the entire temperature and delay-time parameter space. Results are plotted on a *semilog* scale to accentuate the various decay regions, especially the metastability distinctly observed as a ~10 ps *plateau* in the 4 K and 100 K transient signals.[13] This plateau is less apparent in the 200 K and 300 K transients. By 300 K, the hole bands are expected to have flattened sufficiently throughout the Brillouin zone partially, overcoming the indirect nature of the hh3→e1 transition that is believed to be responsible for the metastability.[14] As with the individual AC photoconductivity spectra, the higher temperature data shows a slower decay due to an extended hot-carrier density that is slow to recombine because the charge-to-thermal energy pathway is hindered, both by a phonon bottleneck in the decay of optical-to-acoustic phonon scattering and a low thermal conductivity,[10,11] resulting in reabsorption of optical phonons by the carriers that prevents their thermalization. Therefore, at higher temperatures the higher conductivity persists for longer and is not enhanced or diminished by the lack of the metastability exhibited at lower lattice temperatures.

Scattering times ($\tau$) determined from the plasmon model are related to the carrier mobility $\mu = e\tau/m^*$. Since $\tau$ are simultaneously found with the $N_{exc}$, $\mu(N_{exc})$ can be plotted to illustrate the effect of carrier-carrier interaction on the mobility for different excitation conditions. Figure 4 (a) shows $\mu(N_{exc})$ for 1.03 eV and 0.93 eV photoexcitation at $T_l = 4$ K. Data points are obtained from both the transient and plasmon-model estimates of $N_{exc}$. At both excitation energies, $\mu(N_{exc})$ starts at similar values and then decrease with higher carrier concentration in a manner that looks like the hyperbolic tangent of a Fermi-Dirac distribution. This is consistent with the doping-concentration dependence of mobility, empirically identified by Caughey and Thomas as sigmoidal curves[32] and applied in photoexcited carrier results for GaAs[33] and Si[34] probed by time-resolved THz spectroscopy. The $\mu(N_{exc})$ data can be empirical fit by

$$\mu^\| = \frac{\mu_{max}^\| - \mu_{min}^\|}{1 + (N_{exc}/N_c)^\beta} + \mu_{min}^\|, \quad (3)$$

where $\mu_{max\ (min)}^\|$ is the maximum(minimum) mobility driven parallel to the THz electric field, and $N_c$ and $\beta$ are the center and slope of the inflection between the maximum and minimum values. For $N_{exc} < 2 \times 10^{12}$ cm$^{-2}$, $\mu_{max}^\| \approx 2.48 \times 10^3$ cm$^2$/Vs and $N_c \approx 1.04 \times 10^{13}$ cm$^{-2}$ for both excitations, whereas $\mu_{min}^\| \approx 2.13(1.85) \times 10^3$ cm$^2$/Vs for 0.93(1.03) eV. The lower $\mu_{min}^\|$ for 1.03 eV excitation is due to an increased carrier-carrier scattering that decreases the mean-free path between scattering events, and hence a decreased drift velocity ($v_d = -\mu^\| E$). Given a maximum driving THz field of ≈25 same approach, namely by fitting the plasmon model to the AC photoconductivity spectra for the MQW sample with excitation at 1.03 eV. Figure 3 shows typical results for $\sigma_r$ and $\sigma_i$ comparing $\Delta t = 3$ ps and 90 ps for (a) $T_l = 4$ K and (b) $T_l = 300$ K. At both temperatures the conductivity at the earlier time is stronger and approximately equal. As with the excitation energy comparison, the excitation density deceases with increasing $\Delta t$, with a reduction in $\sigma_r$(@1 THz) from 3 ps to 90 ps of ≈60% at $T_l = 4$ K and only ≈35% at 300 K. These reductions are consistent with a slower carrier density decay at the higher temperature as the system transition to quasi-type-II alignment as the holes delocalize.[13] All spectra exhibit the characteristics of the plasmon model, so values of kV/cm the maximum $v_d$ is $6.2 \times 10^7$ cm/s for both excitation energies, which drops to minima of $5.33(4.63) \times 10^7$ cm/s for 0.93(1.03) eV excitation. These values are consistent with $v_d$ reported in 2D structures.[35–38]

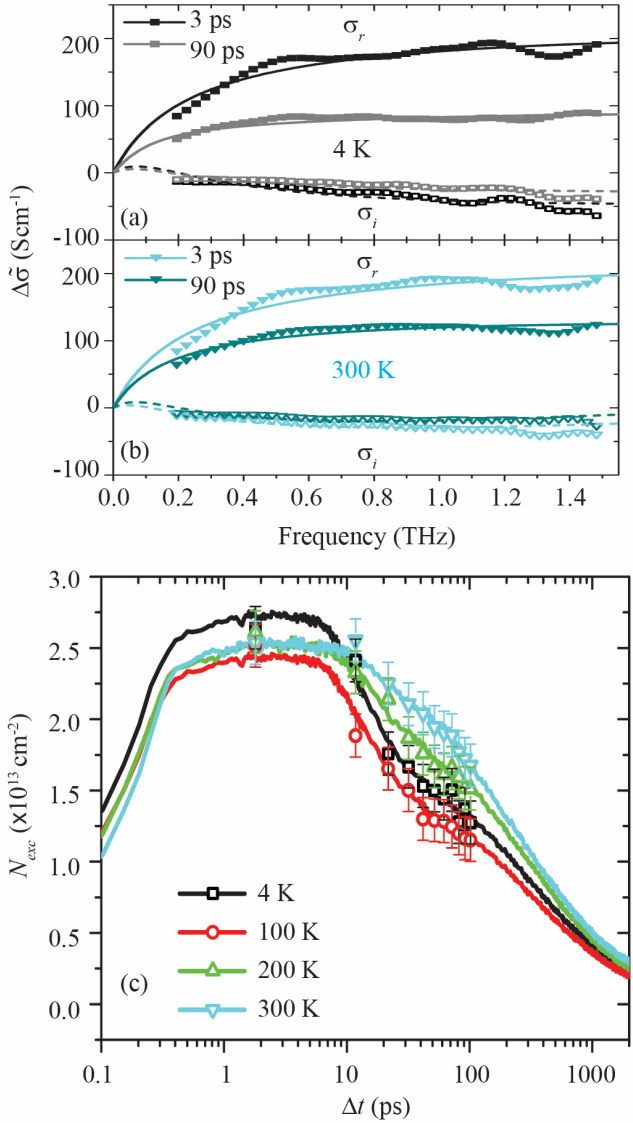

**Fig. 3** AC photoconductivity spectra for the MQW sample excited at 1.03 eV for $\Delta t = 3$ ps and 90 ps at (a) 4 K and (b) 300 K. (c) Photocarrier density decays for excitation of 1.03 eV for a range of lattice temperatures from transient absorption (solid lines) and plasmon modeling (data points).

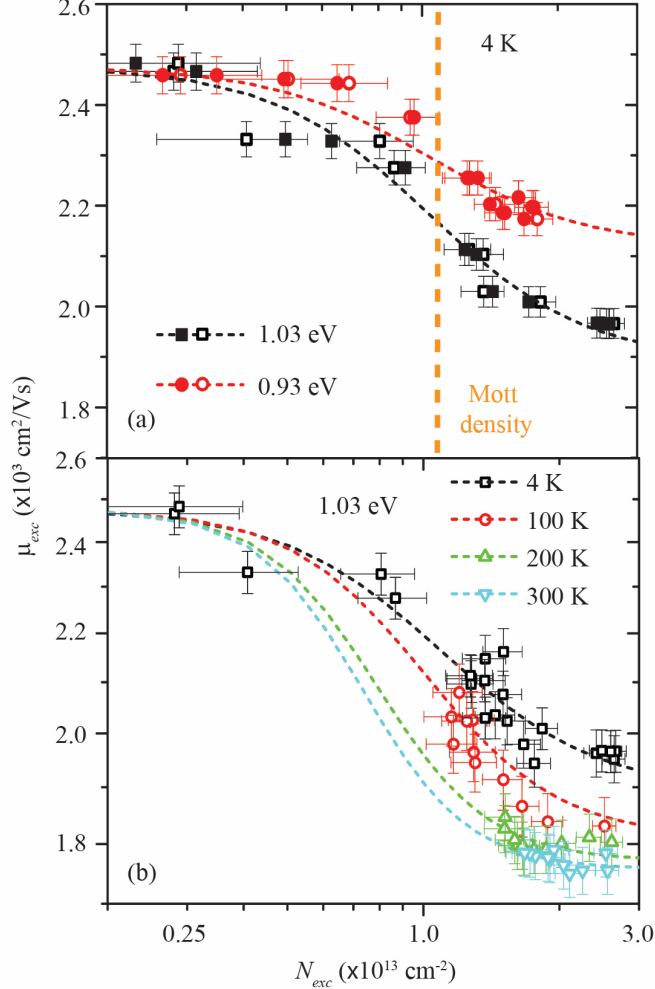

**Fig. 4** Mobility versus carrier density for the QW sample at (a) 4 K, photoexcited at 0.93 eV and 1.03 eV, and (b) photoexcited at 1.03 eV for a range of lattice temperatures. Filled and open symbols represent $\mu(N_{exc})$ extracted carrier density from the transient absorption and the plasmon model respectively. Data are empirically fit with the Caughey-Thomas curves.

Figure 4(b) shows $\mu(N_{exc})$ as a function of $T_l$ for 1.03 eV photoexcitation. For $T_l = 4$ K the data is the same as in Figure 4(a), while the other temperature data are acquired using $\mu$ from the plasmon model and $N_{exc}$ from the transient absorption. Since the increase in photoexcited carrier density can be controlled through the irradiance of the optical pump pulse in much the same way that doping concentration (or back-gate voltage) controls the carrier density, all four data sets are expected to be qualitatively consistent with the empirical Caughey-Thomas curves; fitting values are given in Table 1. For all four temperatures, the maximum mobility is set to $\mu^{\parallel}_{max} \approx 2.48 \times 10^3$ cm²/Vs consistent with the data from Fig 4 (a), because this value is determined from the low-excitation-density region of the transient, where $N_{exc}$ is decaying toward zero regardless of the temperature. Moreover, this value is two orders of magnitude larger than the ground-state mobility (determined without the use of an optical excitation pulse) and is extremely weakly dependent on the lattice temperature.

The remainder of the fitting parameters in Table 1 are temperature dependent. Increasing $T_l$ decreases $\mu^{\parallel}_{min}$ and $N_c$, and increases $\beta$. The reduction in $\mu^{\parallel}_{min}$ may suggest increased scattering at the early times after excitation, even though there is little change in the carrier density. However, unlike the photon energy independence of the position of $N_c$, its change in this case suggests carrier-lattice interactions. The increase in $\beta$ simply means that the transition from $\mu^{\parallel}_{max}$ to $\mu^{\parallel}_{min}$ occurs over a narrower range of carrier density and is not shown in Table 1.

## 4. Discussion

Carrier mobility is related to the MQW structure and it is known that electrodynamics properties of nanostructures is often excitonic, especially when length scales within the structure are smaller than the excitonic Bohr radius.[39,40] In this case, because the AC photoconductivity fits to a plasmon model, excited carriers are considered as electron-hole pairs, that are loosely bound by columbo interactions and experience ambipolar diffusion. The exciton Bohr radius is $a_X = \hbar^2 \epsilon / e^2 m_r^* \approx 3$ nm,[41,42] where $\epsilon = \epsilon_r \epsilon_0$ and $m_r^*$ are the permittivity and reduced effective mass estimated from values of InAs and AlAsSb,[43] and $\hbar$ is the reduced Planck constant. From the Bohr radius, the Mott density is estimated to be $N_{Mott} = 1/a_X^2 \approx 1.1 \times 10^{13}$ cm⁻²,[44] which denotes the transition from a gas to a plasma of the charge carriers, and is illustrated as dashed vertical line in Fig. 4(a). For comparison, the recombination dynamics have previously been assessed for various temperatures,[13] demonstrating that at early times of the transient, corresponding to $\mu_{min}^{\parallel}$, are likely dominated by Auger recombination, whereas $N_c(4\,K) \approx N_{Mott}$ occurs in the carrier density range where radiative recombination dominates, and $\mu_{max}^{\parallel}$ corresponds to a carrier density dominated by Shockley-Reed-Hall recombination. For low temperature, the Mott density is slightly below the transition from dynamics dominated by radiative recombination ($N_{exc} > N_{Mott}$) to Auger recombination $N_{exc} < N_{Mott}$. For the low temperature, it is reasonable that Auger scattering leads to break up of electron-hole pairs and creates a plasma of the majority carriers. At higher temperatures, this transition to Auger recombination does not occur, so while the dynamics offer insight into low temperature behavior, they do not explain the mobility completely.

*Table 1.* Caughey-Thomas curve fitting parameters, calculated ambipolar diffusion coefficient and mean free path for a range of lattice temperatures for 1.03 eV excitation.

| $T_l$ (K) | $\mu^{\parallel}_{min}$ ($\times 10^3$ cm²/Vs) | $N_c$ ($\times 10^{13}$ cm⁻²) | $D_{am}$ (cm²/s) | $L_D$ (nm) | $l_f$ (nm) |
|---|---|---|---|---|---|
| 4   | 1.85 ± 0.08 | 1.04 ± 0.05 | 0.4 ± 0.03   | 24.5 ± 0.5  | 2.13 ± 0.07  |
| 100 | 1.80 ± 0.08 | 0.95 ± 0.05 | 10.19 ± 0.67 | 174.8 ± 0.9 | 9.92 ± 0.38  |
| 200 | 1.78 ± 0.08 | 0.75 ± 0.04 | 20.44 ± 1.34 | 391.5 ± 2.7 | 13.82 ± 0.54 |
| 300 | 1.76 ± 0.08 | 0.70 ± 0.04 | 30.74 ± 2.01 | 619.9 ± 3.5 | 16.67 ± 0.66 |

A drift-diffusion model can be applied to the transport of the native and excited carriers. Shortly after excitation, photocarriers act as mostly neutral electron-hole pairs, so that they are only partly driven by the probe THz electric field. Moreover, the drift contribution is weakly dependent on temperature because the THz probe is fixed in strength. Hence, the following analysis is limited to ambipolar diffusion[45] determined by an Einstein model,

$$D_{am}(T_l) = \frac{k_B T_l}{e} \left[ \frac{(2N_{exc} + P_0)}{\left(\frac{N_{exc}}{\mu_p} + \frac{N_{exc} + N_0}{\mu_n}\right)} \right], \quad (4)$$

where $N_0$ is native carrier density estimated from ground-state AC conductivity and photoluminescence to be $\approx 0.5(1.5) \times 10^{11}$ cm$^{-2}$ at low(high) $T_l$, and $\mu_{n(p)}$ are electron(hole) mobility estimated from $\mu_{min}^{\parallel} \approx \mu_n \mu_p / (\mu_n + \mu_p)$ because $N_{exc} \gg N_0$.[45]

In Table 1, $D_{am}$ is shown to increase as a function of temperature, which results in an increased in the average distance a carrier travels between excitation and recombination. This is characterized by the diffusion length $L_D = \sqrt{D_{am} \tau_R}$,[46,47] where $\tau_R$ is the carrier lifetime. Table 1 also shows $L_D$ values estimated with $\tau_R$ set as the first decay component from fitting the transient absorption.[13] Figure 3(c) shows that $\tau_R$ increases with lattice temperature and leads to a larger $N_{exc}$ at longer $\Delta t$ delays. Hence, as $T_l$ increases, $\tau_R$ and $\mu_{min}^{\parallel}$ are inversely correlated and the measured values $\mu_{min}^{\parallel}$ persist to lower values of $N_{exc}$. This effect results in a decreasing $N_c$ with $T_l$.

Overall, as the carrier dynamics in the region of $\mu_{min}^{\parallel}$ transition from Auger to radiative recombination with increasing temperature, $D_{am}$ and $L_D$ increase, and $\mu_{min}^{\parallel}$ decreases as is seen in the Fig. 4(b). Increased diffusion might be expected to spread the carriers out in real space and reduce the likelihood of Auger scattering. Additionally, this can be related to the mean-free path $l_f$ that is also tabulated. Comparison shows that both $L_D$ and $l_f$ increase with $T_l$, but the ratio $L_D/l_f$ reveals that there are approximately 11(37) scattering events on average before recombination at low(high) $T_l$. Figure 5 (a)

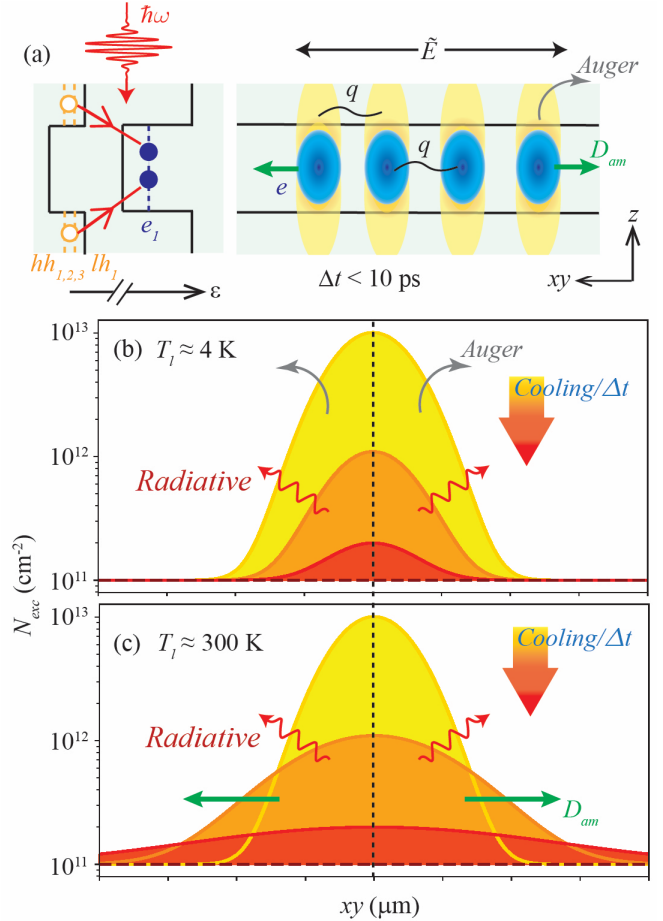

**Fig. 5** (a) Schematic diagram of the type-II band alignment, photoexcitation of charge carriers into $e_1$, $hh_{1-3}$, $lh_1$ bands, and the interaction of carriers including diffusion ($D_{am}$), Auger scattering and electron-phonon scattering ($q$). [$\varepsilon$ is band energy, $\hbar\omega$ is the photon energy.] These interactions vary the spatial distribution of the cooling charge carriers depending on the lattice temperature and are illustrated for (b) $T_l \approx 4$ K and (c) $T_l \approx 300$ K.

summarizes the photocarrier dynamics (Auger, then radiative then Shockley-Read-Hall) and transport through diffusion. Schematic diagrams of the photocarrier distribution during the cooling process are shown at (b) $T_l \approx 4$ K, where strong Auger recombination initially occurs, and (c) $T_l \approx 300$ K, where diffusion quickly spreads the distribution. The increased scattering at higher temperature is consistent with a decreased $\mu_{min}^{\parallel}$. Hence, in the extremely high excitation regime, lower temperature leads to both the metastable dynamics and the better mobility.

Finally, the sample temperature affects the photocarrier transport through interactions with the lattice. The data shows that $\mu_{max}^{\parallel}(T_l)$ is temperature independent and that $\mu_{min}^{\parallel}(T_l)$ decreases somewhat linearly by only $\approx 5\%$ over the temperature range. The weakness of the latter temperature dependence does not match any individual component of the ground-state carrier-lattice interaction which are determined in either THz time-domain spectroscopy or Hall-effect measurements.[48]

Namely, these are carrier mobility $\propto T_l^{-0.5}$ due to deformation potential scattering, $\propto T_l^{1.5}$ due to piezoelectric potential scattering and $\propto \exp(1/T_l)$ for polar-optical scattering. To maintain an almost $T_l$-independent response, piezoelectric potential scattering by acoustic phonons and polar optical scattering limit the carrier mobility in the low and high temperature regimes respectively. While more data is required to properly interpret the carrier-lattice interactions, it is clear that the carrier density has a more significant effect on the transport than the temperature dependence. This suggests that temperature would play a smaller role in limiting the operational range for InAs/AlAsSb-based devices.

## 5. Conclusion

In summary, AC photoconductivity measurements are presented for a type-II InAs/AlAsSb heterostructures to determine the transport of photocarriers for a range of lattice temperatures and excitation conditions. The results complement photocarrier dynamics that previously identified the recombination mechanisms, and highlighted an intriguing metastability at low temperatures and for sufficient excess excitation energy.[13] While the photon energy dependence shows that carrier transport in the metastable regimes exhibits higher conductivity, the mobility is slightly reduced compared to other excitation regimes.

At lower excitation densities, that are within the reach of moderate solar concentration, photocarrier transport is temperature independent. Additionally, at warmer temperatures close to ambient operating conditions, hot-carrier lifetimes remain prolonged, and ambipolar diffusion and radiative recombination govern the transport and dynamics. Moreover, mobility for low excitation density is comparable to elemental and III-V based photovoltaic devices,[49–51] and exceeds that of organic[52] and hybrid-perovskite[53] solar cells by several orders of magnitude.

Long hot-carrier lifetimes and good mobility can be exploited in novel type-II InAs/AlAsSb device architectures. Furthermore, these systems can further increase hot-carrier lifetimes with processes like intervalley scattering[16] that don't requiring excessive excitation densities.

**Acknowledgements:** The authors wish to thank Oklahoma Photovoltaics Research Institute (OKPVRI) and the Center for Quantum Research and Technology (CQRT) at the University of Oklahoma.

**Funding:** This work was supported by National Science Foundation, CBET-2102239 at West Virginia University and ECCS-1610062 at the University of Oklahoma.

**References**
[1] D.K. Ferry, Semicond. Sci. Technol. **34**, 044001 (2019).
[2] L.C. Hirst and N.J. Ekins-Daukes, Prog. in Photovolt. **19**, 286 (2011).
[3] D.K. Ferry, S.M. Goodnick, V.R. Whiteside, and I.R. Sellers, J. Appl. Phys. **128**, 220903 (2020).
[4] W. Shockley and H.J. Queisser, J. Appl. Phys. **32**, 510 (1961).
[5] T. Kita, Y. Harada, and S. Asahi, in *Energy Conversion Efficiency of Solar Cells*, edited by T. Kita, Y. Harada, and S. Asahi (Springer, Singapore, 2019), pp. 55–79.
[6] D. König, K. Casalenuovo, Y. Takeda, G. Conibeer, J.F. Guillemoles, R. Patterson, L.M. Huang, and M.A. Green, Physica E **42**, 2862 (2010).
[7] S. Kahmann and M.A. Loi, J. Mater. Chem. C 16 (2019).
[8] L.C. Hirst, M.P. Lumb, R. Hoheisel, C.G. Bailey, S.P. Philipps, A.W. Bett, and R.J. Walters, Sol. Energy Mater. Sol. Cells **120**, 610 (2014).
[9] H. Esmaielpour, V.R. Whiteside, J. Tang, S. Vijeyaragunathan, T.D. Mishima, S. Cairns, M.B. Santos, B. Wang, and I.R. Sellers, Prog. Photovolt: Res. Appl. **24**, 591 (2016).
[10] H. Esmaielpour, V.R. Whiteside, H.P. Piyathilaka, S. Vijeyaragunathan, B. Wang, E. Adcock-Smith, K.P. Roberts, T.D. Mishima, M.B. Santos, A.D. Bristow, and I.R. Sellers, Sci. Rep. **8**, 12473 (2018).
[11] H. Esmaielpour, B.K. Durant, K.R. Dorman, V.R. Whiteside, J. Garg, T.D. Mishima, M.B. Santos, I.R. Sellers, J.-F. Guillemoles, and D. Suchet, Appl. Phys. Lett. **118**, 213902 (2021).
[12] J. Tang, V.R. Whiteside, H. Esmaielpour, S. Vijeyaragunathan, T.D. Mishima, M.B. Santos, and I.R. Sellers, Appl. Phys. Lett. **106**, 061902 (2015).
[13] H.P. Piyathilaka, R. Sooriyagoda, H. Esmaielpour, V.R. Whiteside, T.D. Mishima, M.B. Santos, I.R. Sellers, and A.D. Bristow, Sci. Rep. **11**, 10483 (2021).
[14] V.R. Whiteside, B.A. Magill, M.P. Lumb, H. Esmaielpour, M.A. Meeker, R.R.H.H. Mudiyanselage, A. Messager, S. Vijeyaragunathan, T.D. Mishima, M.B. Santos, I. Vurgaftman, G.A. Khodaparast, and I.R. Sellers, Semicond. Sci. Technol. **34**, 025005 (2019).
[15] V.R. Whiteside, H. Esmaielpour, T.D. Mishima, K.R. Dorman, M.B. Santos, D.K. Ferry, and I.R. Sellers, Semicond. Sci. Technol. **34**, 094001 (2019).
[16] H. Esmaielpour, K.R. Dorman, D.K. Ferry, T.D. Mishima, M.B. Santos, V.R. Whiteside, and I.R. Sellers, Nat. Energy **5**, 336 (2020).
[17] R. Dixit, B. Barut, S. Yin, J. Nathawat, M. Randle, N. Arabchigavkani, K. He, C.-P. Kwan, T.D. Mishima, M.B. Santos, D.K. Ferry, I.R. Sellers, and J.P. Bird, Phys. Rev. Materials **4**, 085404 (2020).
[18] D. Mittleman, J. Cunningham, M. Nuss, and M. Geva, Appl. Phys. Lett. **71**, 16 (1997).
[19] H.J. Joyce, J.L. Boland, C.L. Davies, S.A. Baig, and M.B. Johnston, Semicond. Sci. Technol. **31**, 103003 (2016).
[20] J. Lloyd-Hughes and T.-I. Jeon, J Infrared Milli Terahz Waves **33**, 871 (2012).
[21] H.P. Piyathilaka, R. Sooriyagoda, V. Dewasurendra, M.B. Johnson, K.T. Zawilski, P.G. Schunemann, and A.D. Bristow, Opt. Express **27**, 16958 (2019).
[22] J.D. Rowley, J.K. Wahlstrand, K.T. Zawilski, P.G. Schunemann, N.C. Giles, and A.D. Bristow, Opt. Express **20**, 16968 (2012).

[23] H. Kroemer, Physica E **20**, 196 (2004).
[24] Q. Wu, M. Litz, and X.-C. Zhang, Appl. Phys. Lett. **68**, 2924 (1996).
[25] P.C.M. Planken, H.-K. Nienhuys, H.J. Bakker, and T. Wenckebach, J. Opt. Soc. Am. B **18**, 313 (2001).
[26] C. Winnewisser, P.U. Jepsen, M. Schall, V. Schyja, and H. Helm, Appl. Phys. Lett. **70**, 3069 (1997).
[27] D. Grischkowsky, S. Keiding, M. van Exter, and C. Fattinger, J. Opt. Soc. Am. B **7**, 2006 (1990).
[28] E. Mavrona, F. Appugliese, J. Andberger, J. Keller, M. Franckié, G. Scalari, and J. Faist, Opt. Express **27**, 14536 (2019).
[29] J. Linnros, J. Appl. Phys. **84**, 275 (1998).
[30] T.R. Senty, S.K. Cushing, C. Wang, C. Matranga, and A.D. Bristow, J. Phys. Chem. C **119**, 6337 (2015).
[31] R. Ulbricht, E. Hendry, J. Shan, T.F. Heinz, and M. Bonn, Rev. Mod. Phys. **83**, 543 (2011).
[32] D.M. Caughey and R.E. Thomas, Proc. IEEE **55**, 2192 (1967).
[33] M.C. Beard, G.M. Turner, and C.A. Schmuttenmaer, Phys. Rev. B **62**, 15764 (2000).
[34] T.J. Magnanelli and E.J. Heilweil, Opt. Express **28**, 7221 (2020).
[35] V.G. Mokerov, I.S. Vasil'evskii, G.B. Galiev, J. Požela, K. Požela, A. Sužiedėlis, V. Jucienė, and Č. Paškevič, Semicond. **43**, 458 (2009).
[36] A. Guen-Bouazza, C. Sayah, B. Bouazza, and N.E. Chabane-Sari, J. Mod. Phys. **04**, 616 (2013).
[37] J.S. Blakemore, J. Appl. Phys. **53**, R123 (1982).
[38] A. Šilenas, Yu. Požela, K. Požela, V. Jucienė, I.S. Vasil'evskii, G.B. Galiev, S.S. Pushkarev, and E.A. Klimov, Semicond. **47**, 372 (2013).
[39] S. Wu, L. Cheng, and Q. Wang, Mater. Res. Express **4**, 085017 (2017).
[40] B. Jiang, C. Zhang, X. Wang, F. Xue, M.J. Park, J.S. Kwak, and M. Xiao, Opt. Express **20**, 13478 (2012).
[41] A. Hangleiter, Z. Jin, M. Gerhard, D. Kalincev, T. Langer, H. Bremers, U. Rossow, M. Koch, M. Bonn, and D. Turchinovich, Phys. Rev. B **92**, 241305 (2015).
[42] B. Gerlach, J. Wüsthoff, M.O. Dzero, and M.A. Smondyrev, Phys. Rev. B **58**, 10568 (1998).
[43] I. Vurgaftman, J.R. Meyer, and L.R. Ram-Mohan, J. Appl. Phys. **89**, 5815 (2001).
[44] N.F. Mott, Rev. Mod. Phys. **40**, 677 (1968).
[45] H. Hempel, C.J. Hages, R. Eichberger, I. Repins, and T. Unold, Sci. Rep. **8**, (2018).
[46] A. Fiore, M. Rossetti, B. Alloing, C. Paranthoen, J.X. Chen, L. Geelhaar, and H. Riechert, Phys. Rev. B **70**, 205311 (2004).
[47] C.-H. Chu, M.-H. Mao, Y.-R. Lin, and H.-H. Lin, Sci. Rep. **10**, 5200 (2020).
[48] R. Sooriyagoda, H.P. Piyathilaka, K.T. Zawilski, P.G. Schunemann, and A.D. Bristow, J. Opt. Soc. Am. B **38**, 769 (2021).
[49] H. Cotal, C. Fetzer, J. Boisvert, G. Kinsey, R. King, P. Hebert, H. Yoon, and N. Karam, Energy Environ. Sci. **2**, 174 (2009).
[50] K. Makita, H. Mizuno, T. Tayagaki, T. Aihara, R. Oshima, Y. Shoji, H. Sai, H. Takato, R. Müller, P. Beutel, D. Lackner, J. Benick, M. Hermle, F. Dimroth, and T. Sugaya, Prog. Photovolt. **28**, 16 (2020).
[51] J. Li, A. Aierken, Y. Liu, Y. Zhuang, X. Yang, J.H. Mo, R.K. Fan, Q.Y. Chen, S.Y. Zhang, Y.M. Huang, and Q. Zhang, Front. Phys. **8**, 631925 (2021).
[52] A. Melianas, V. Pranculis, Y. Xia, N. Felekidis, O. Inganäs, V. Gulbinas, and M. Kemerink, Adv. Energy Mater. **7**, 1602143 (2017).
[53] J. Wu, H. Cha, T. Du, Y. Dong, W. Xu, C.-T. Lin, and J.R. Durrant, Adv. Mater. **34**, 2101833 (2022).